\documentclass{emulateapj}
\usepackage{epsfig}

\newcommand{\beq}{\begin{equation}}
\newcommand{\eeq}{\end{equation}}

\def\be{\begin{equation}}
\def\ee{\end{equation}}
\def\bea{\begin{eqnarray}}
\def\eea{\end{eqnarray}}



\def \logTd6 {\hbox{log$( T/6 \kev)$} }

\def\myputfigure#1#2#3#4#5%
{\vskip#5pt\makebox[0pt]{\hskip#2in
\includegraphics[width=#3\textwidth]{#1}}\vskip#4pt\hfill}

\def \arcsec    {^{\prime\prime}}

\def \kms            {~{\rm km~s}^{-1}}

\def \etal      {et al.}

\def \kmsmpc    {{\rm\ km\ s^{-1}\ Mpc^{-1}}}
\def \kev       {{\rm\ keV}}

\def \hmsol     {h^{-1}{\rm\ M}_\odot}
\def \hMpc      {h^{-1}{\rm\ Mpc}}
\def \hkpc      {h^{-1}{\rm\ kpc}}

\input epsf
\bibpunct{(}{)}{;}{a}{}{,}

\begin{document}

\lefthead{QUASARS LENSED BY CLUSTERS}
\righthead{Hennawi \etal}

\author{Joseph F. Hennawi\altaffilmark{1,2,3}, 
  Neal Dalal\altaffilmark{1,4}
  Paul Bode\altaffilmark{3}
} 

\altaffiltext{1}{Hubble Fellow}

\altaffiltext{2}{Department of Astronomy, University of California
  Berkeley, Berkeley, CA 94720}

\altaffiltext{3}{Princeton University Observatory, Princeton, NJ 08544}

\altaffiltext{4}{Institute for Advanced Study, Einstein Drive,
  Princeton, NJ 08540}

\title{Statistics of Quasars Multiply Imaged by Galaxy Clusters}
\label{chap:qsolens}
\begin{abstract}
We compute the expected number of quasars multiply imaged by cluster
size dark halos for current wide field quasar surveys by carrying out
a large ensemble of ray tracing simulations through clusters from a
cosmological N-body simulation of the $\Lambda$CDM cosmology.  Our
calculation predicts $\sim 4$ quasar lenses with splittings
$\theta>10\arcsec$ in the SDSS spectroscopic quasar sample, consistent
with the recent discovery of the wide separation lens SDSSJ1004+4112
which has $\theta=14\arcsec.6$. The SDSS faint photometric quasar
survey will contain $\sim 12$ multiply imaged quasars with splittings
$\theta>10\arcsec$. Of these, $\sim 2$ will be lenses with
separations $\theta>30\arcsec$, and $\sim 2$ will be at high redshift ($z_{\rm
  s}\sim 4$).
 \end{abstract}

\section{Introduction}
\label{sec:intro}

Strong gravitational lensing by clusters of galaxies provides a unique
laboratory for studying the dark matter distribution in the largest
collapsed structures in the universe.  The most common manifestation
of cluster strong lensing are giant arcs, which are highly elongated
and sometimes multiply imaged background galaxies which have been
distorted by the deep gravitational potential of a massive galaxy
cluster.  Background quasars can also be multiply imaged by foreground
lenses. However, until recently, all of the roughly 70 lensed quasars
known \citep{Castles} had small $\theta\lesssim 10\arcsec$
separations, explained in terms of galaxy-scale concentrations of
baryonic matter, despite a number of searches for lenses with wider
splittings \citep{Maoz97,Ofek01,Phillips01,ZS01,Miller04,Marble04}.



The discovery of the quadruply imaged quasar SDSSJ1004+4112
\citep{Quad03,Quad04,Inada05} in the Sloan Digital Sky Survey (SDSS), with a
largest image separation of $\theta=14\arcsec.6$, constitutes the
first quasar multiply imaged by a cluster where dark matter dominates
the lensing potential \citep{Quad04}. Quasars lensed
by clusters are much rarer than giant arcs simply because the number
density of quasars behind clusters is much lower than that of
galaxies.  Because massive clusters capable of strong lensing are also
very rare, large samples of quasars are required to find quasars 
multiply imaged by clusters.

The advent of large area digital imaging and spectroscopic surveys,
like the SDSS \citep{York00} and the Two Degree Field Quasar Survey
\citep{Croom04}, has dramatically increased the number of known
quasars. Although a spectroscopic survey of $\sim$ 30,000 quasars was
required to discover one multiply imaged quasar lensed by a cluster
\citep{Quad03,Quad04}, new \emph{photometric} quasar selection
techniques applied to the current SDSS imaging data promise to yield
as many as a million quasars \citep{Richards04}.  In addition, future
large area synoptic surveys such as Pan-STARRS and the LSST will be
able to use variability selection techniques to detect as many as one
hundred million quasars.  The number of quasars lensed by clusters
discovered in these current and future samples of quasars will allow
statistical studies of the abundance of these large separation quasar
lenses.
 
It has been suggested that the abundance of cluster lenses can be used
as a cosmological probe. \citet{Bart98} found that the predicted
number of giant arcs varies by orders of magnitude among different
cosmological models, and \citet{Bart03} and \citet{Mene04} recently
proposed that giant arc statistics could be used to distinguish
between different dark energy cosmological models. Despite the fact
that the abundance of quasars lensed by clusters will be significantly
lower, because quasars are point sources, the selection function of
multiply imaged quasars is much easier to quantify than that of giant
arcs, because the ability to detect giant arcs is extremely sensitive
to the seeing.  Indeed, beginning with \citet{Koch96}, the abundance
of multiply imaged quasars has been recognized as a potentially
powerful cosmological probe. However, the dominant uncertainty has
always been the accuracy with which the galaxy population can be
modeled. Because on cluster scales only dark matter and gravity are
involved, quasars multiply imaged by clusters would not suffer from
these modeling uncertainties.

The measurement of a time delay in multiply imaged quasar cluster lens
could prove to be promising method for measuring the Hubble
constant, if the cluster lens \emph{also shows giant arcs}. This is
because there would be more constraints on the mass model than are
present in galaxy lenses, and it precisely this modeling uncertainty
which limits the precision of this technique \citep[see e.g.][]{Koch03,KS04}. 

Finally, because multiply imaged quasar lenses are found by searching for pairs
of quasars at the same redshift (that is, by surveying the source population), 
whereas giant arcs are discovered by looking for elongated structures in the
vicinity of rich galaxy clusters (that is, by surveying the lens population), 
the two types of cluster lens surveys are complementary.

Nearly all previous investigations of the abundance of quasars lensed
by massive dark halos have used simple spherically symmetric
analytical profiles for the mass distribution of clusters to predict
the abundance of wide separation quasar lenses
\citep{Maoz97,SRM01,KM01,WTS01,LO02,Rusin02,LO03,Oguri03,HM04,
  LM04,KKM04,Quad04,Chen04}. However, in the context of giant arc
statistics, full ray tracing computations through clusters from N-body
simulations have unequivocally demonstrated that analytical models
fail to accurately represent strong lensing by clusters \citep{BSW95},
underestimating the lensing cross sections by as much as two orders of
magnitude \citep{Mene03a,Hennawi05a}.  In recent progress, \citet{OK04}
computed the abundance of quasars multiply imaged by clusters
analytically, using triaxial rather than spherical halos; they found
that triaxiality significantly increases the expected number of
lenses. In addition, all previous predictions for the number of wide
separation multiply imaged quasar lenses have neglected the effect of
the brightest cluster galaxies on the strong lensing cross section,
focusing only on the effects of the dark matter. However, brightest
cluster galaxies can enhance strong lensing cross sections by up to
$\sim 50\%$ \citep{Mene03b,DHH04,Shirley04,Hennawi05a} for small image
separations ($\theta \lesssim 20\arcsec$).



In this paper we compute the expected number of quasars
multiply imaged by cluster size dark halos in current wide field
quasar surveys by carrying out a large ensemble of ray tracing
simulations through clusters from a large cosmological N-body
simulation.  In \S~\ref{sec:Nbody}, we briefly summarize the ray
tracing simulations and our technique for adding BCGs to dark matter
halos.  The details of the quasar luminosity functions for the surveys
we consider are discussed in \S \ref{sec:qso_lum}.  We present results
on the abundance of quasars lensed by clusters in
\S~\ref{sec:lens}, and we conclude in \S \ref{sec:conc}.

\section{Ray Tracing Simulations}
\label{sec:Nbody}

\begin{figure*}
  \centerline{
    \epsfig{file=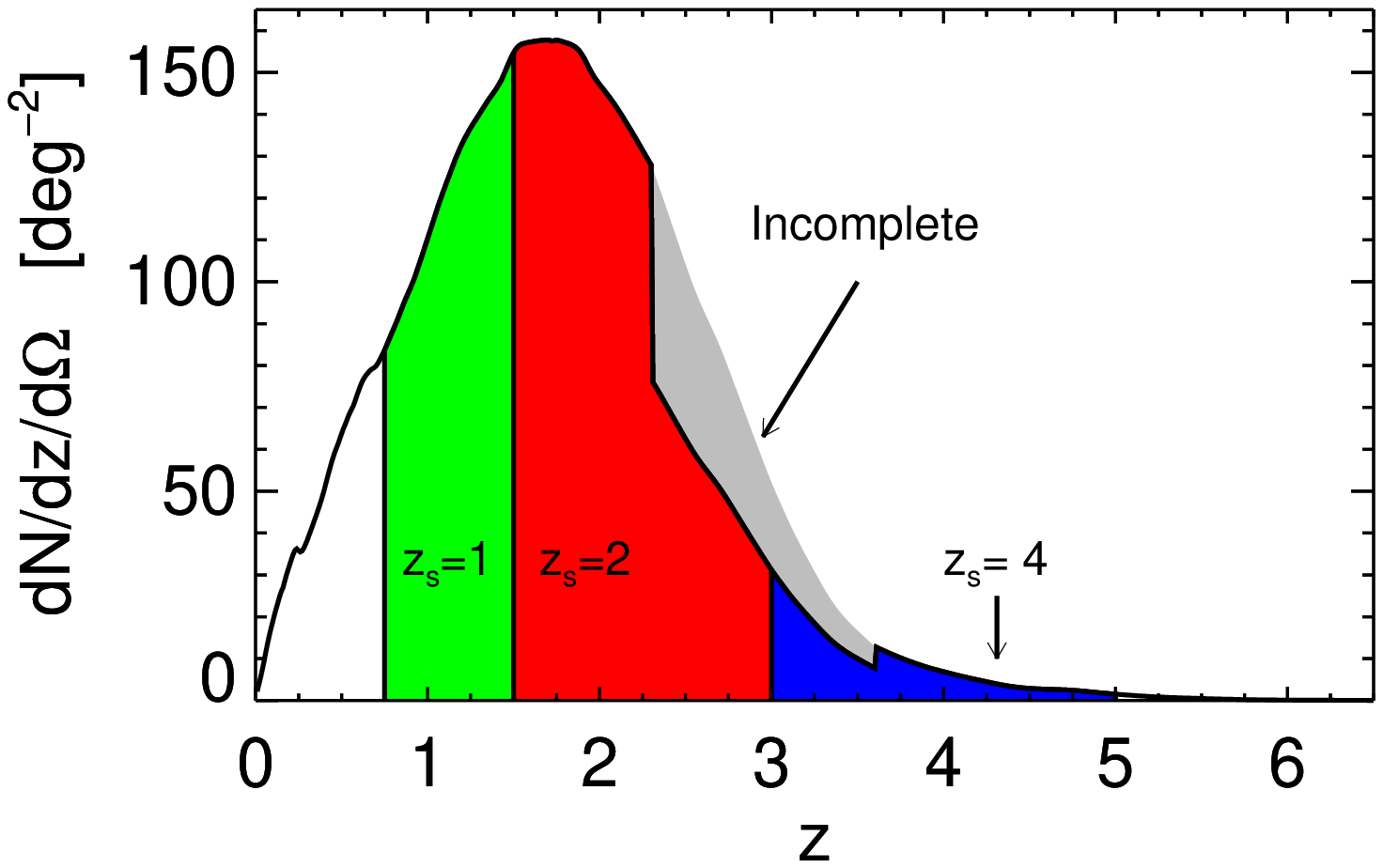,bb=10 10 445 300,width=0.50\textwidth}
    \epsfig{file=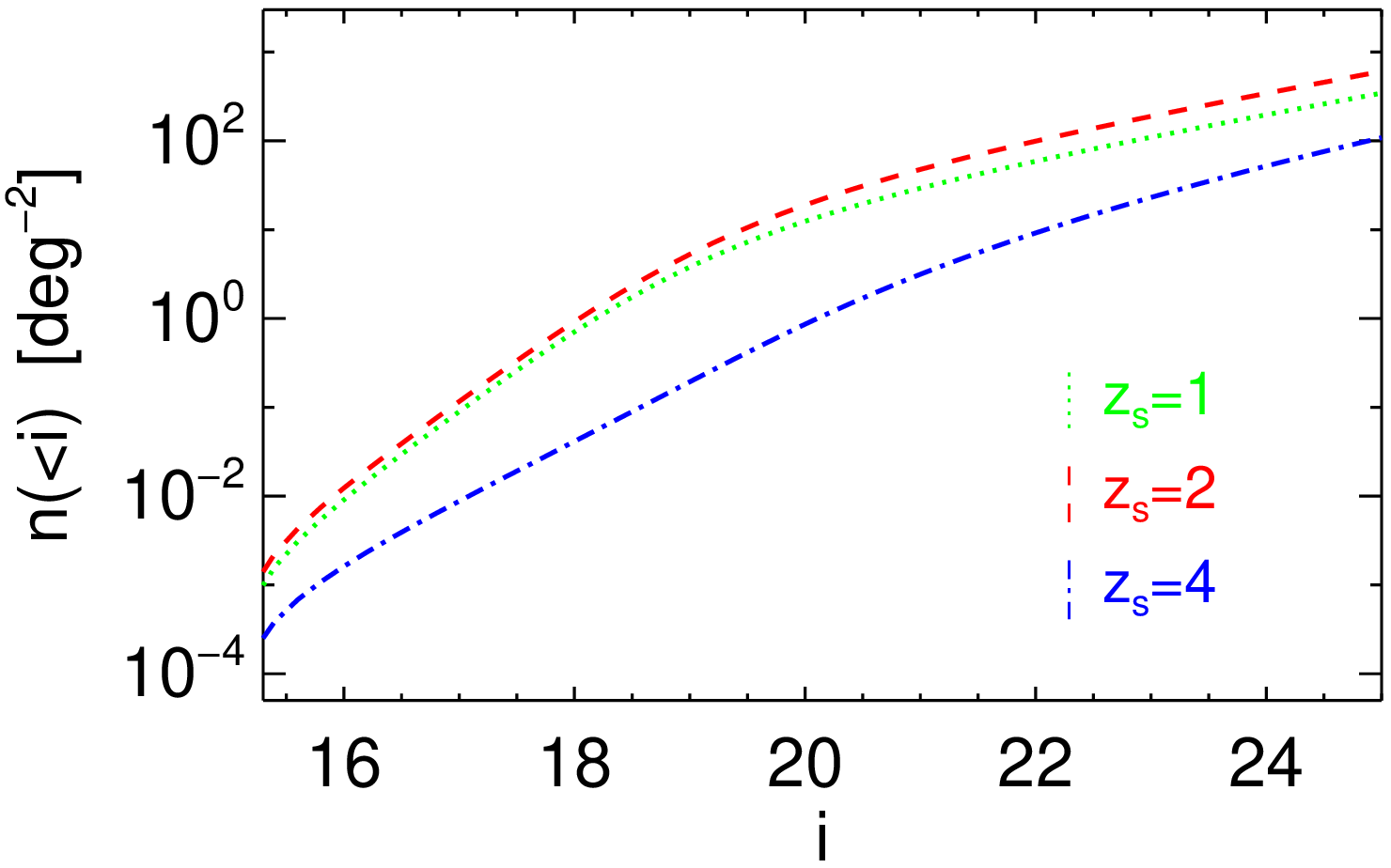,bb=10 10 445 300,width=0.50\textwidth}}
  \caption{ \emph{Left:} Redshift distribution of quasars for a flux
    limit of $i<22.7$. The green, red, and blue shaded regions
    indicate the redshift bins used to determine the number of quasars
    for each of the three source planes at $z_{\rm s}=1$, $2$, and
    $4$, respectively. The area shaded gray represents the number of
    quasars excluded by our assumption of $60\%$ completeness over the
    redshift range $2.3<z<3.6$, where color selected samples are
    incomplete because quasar colors cross the stellar
    locus. \emph{Right:} Cumulative number magnitude counts of quasars
    used for each of our three source planes. These curves were
    obtained by integrating the redshift distribution of quasars over
    redshift bins illustrated by the shading in the left panel.
    \label{fig:qso_lum}
  }
\end{figure*}

In this section,
we briefly summarize the essential elements of our strong lensing
simulations.  We refer the reader to \citet{DHH04} and
\citet{Hennawi05a} for details. We will discuss the N-body simulations, ray
tracing simulations, and our method for adding brightest cluster
galaxies in turn.

In order to simulate the strong lensing effects of a Universe filled
with dark matter, we used clusters from a cosmological N-body
simulation, performed with the Tree-Particle-Mesh (TPM) code of
\citet{BO03}.  The following cosmological parameters were used for
this simulation: $\Omega_{\rm m}=0.3$,
$\Omega_\Lambda=0.7$,$h=70~\kmsmpc$, $\sigma_8$=0.95, and $n_s$=1,
which are consistent (within 1$\sigma$) of the WMAP derived
cosmological parameters \citep{Spergel03}.  The simulations were
performed in a box with a comoving side length of $L=320~\hMpc$ and
$N=1024^3$ particles, giving a particle mass of $m_{\rm p}=2.54\times
10^9 \hmsol$.  The cubic spline softening length was set to
$\epsilon=3.2 \hkpc$.  We use outputs at seven different `snapshots',
covering the range of redshifts over which the critical density is low
enough to produce an appreciable amount of strong lensing:
$z=0.17,0.29,0.41,0.55,0.70,0.87,1.05,1.26,$ and $1.49$. This same
simulation has been used to investigate the statistics of giant arcs
\citep{Wamb04a,Hennawi05a},  strong lensing cosmography
\citep{DHB05}, and the lensing effects of secondary matter 
in the light cone \citep{Wamb04b}.

A ``friends--of--friends'' (FOF) group finder \citep{Davis85} with the
canonical linking length of $b=0.2$ was applied to each particle
distribution to identify cluster size dark matter halos. For each
cluster with a FOF group mass above $M_{\rm FOF}\geq 10^{14}\hmsol$,
all the particles within a $5\hMpc$ sphere about the center of mass
were dumped to separate files and used as inputs to our ray tracing
code.

To compute reliable mean cross sections for a cluster, we must average
over many projections to appropriately sample the distribution of
cross sections \citep{DHH04,Shirley04,Hennawi05a}. We computed mean
lensing cross sections by averaging over 125 orientations for all
clusters with $M_{\rm FOF}\ge 10^{15.0}~\hmsol$, 31 orientations for
clusters in the mass interval $10^{14.5}~\hmsol\le M_{\rm
  FOF}<10^{15.0}~\hmsol$, and 3 orientations for all clusters in the
range $10^{14}~\hmsol\le M_{\rm FOF}<10^{14.5}~\hmsol$.

The surface density of each cluster projection is computed on a grid
and a deflection angle map is computed using FFT methods.  Using
linked lists, a mapping is constructed between source plane pixels and
image plane pixels.  We then compute the statistics of multiply imaged
quasars using a Monte Carlo approach.  Single pixel sources are
randomly placed in the source plane.  If there is more than one image
plane pixel corresponding to this source plane pixel we count this as
a multiply imaged quasar. If the image multiplicity is larger than
two, the separation of the images is taken to be the largest image
separation in the system, and this is considered a single lensing
event, thus we properly count triples, quads, etc.  The magnifications
of each image are also computed, which is just the ratio of the number
of pixels in each image to the original number in the source (which is
unity).

Given this list of image splittings and magnifications, the number of
multiply imaged quasars behind each cluster orientation can be
computed, given the number counts of background quasars as a function
of apparent magnitude.  Our calculation completely accounts
for the magnification bias, because we use the magnification of each
image to determine the corresponding number density of background sources
for the particular survey under consideration.

The ray trace and Monte Carlo must be repeated for each projection
through the cluster and each source plane considered. We ray trace
three source planes at $z_{\rm s}=1.0$, $2.0$, and $4.0$.  The number
counts of quasars are thus collapsed into three source redshift bins
centered on these source planes (see \S \ref{sec:qso_lum}), given by
$[0.75,1.5]$, $[1.5,3.0]$, and $[3.0,5.0]$, respectively. Because the
critical density for strong lensing is a slowly varying function of
source redshift, this binning will not introduce significant errors in
our calculation.

Brightest cluster galaxies (BCGs) can enhance strong lensing cross
sections by $\sim 50\%$ \citep{Mene03b,DHH04,Shirley04,Hennawi05a} for
multiply imaged quasar separations $\lesssim 20^{\prime\prime}$, small
enough such that the mass enclosed by the critical lines has a
significant baryonic component
\citep{DHH04,Shirley04,Hennawi05a}. Indeed, the recently discovered wide
separation quadruply imaged quasar lens SDSS1004+4112 \citep{Quad03}
is centered on the brightest galaxy of the cluster. Models of the lens
which include the central galaxy embedded in a dark matter halo give
an Einstein radius of $\sim 7^{\prime\prime}$, with the galaxy
component contributing a significant amount to the lensing potential
\citep{Quad04}.

We artificially add baryons, with a singular isothermal sphere
profile, to the centers of each cluster by `painting' BCGs onto the
dark matter surface density.  Because varying the total mass of the
central galaxy has a negligible effect on the lensing cross section
\citep{DHH04}, we use a fixed fraction, $M_{\rm baryon}=0.003~M_{\rm
  FOF}$, of the mass of the dark halo. We use a simple scaling prescription
for assigning velocity dispersions to the BCGs
\be
\frac{\sigma}{300 \kms}=\left(\frac{M_{\rm FOF}}{10^{15} \hmsol}\right)^{2/15},
\ee
which is based on an empirical scaling relation between the optical 
luminosity of BCGs and the X-ray temperatures of galaxy clusters 
\citet{ES91} and the fact that BCGs lie on the fundamental plane of 
elliptical galaxies \citep{OH91}. See \citet{Hennawi05a} for details.

\section{Modeling the Quasar Luminosity Function}
\label{sec:qso_lum}
\begin{figure*}
\centerline{
  \epsfig{file=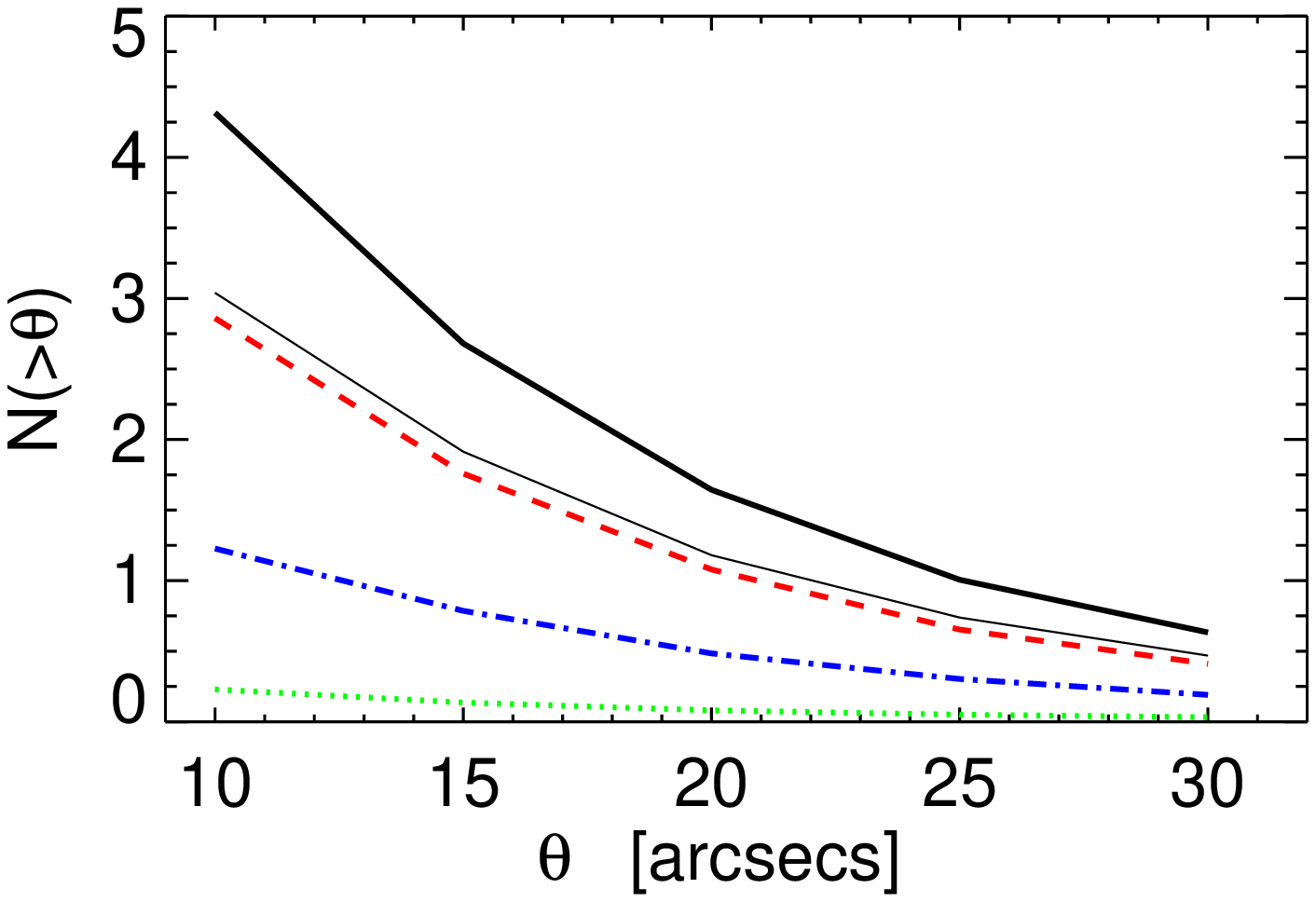,bb=10 10 430 300,width=0.50\textwidth}
  \epsfig{file=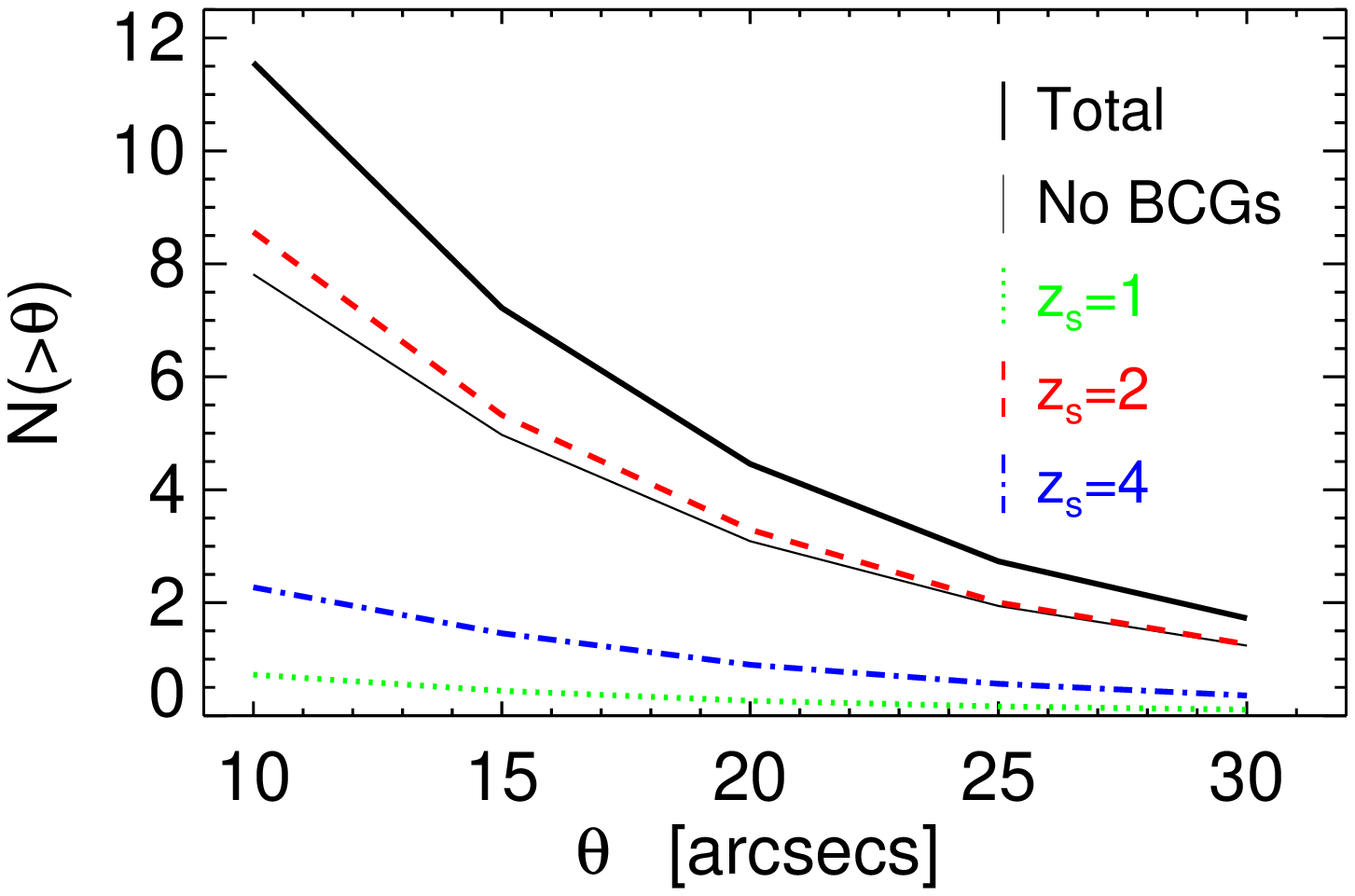,bb=10 10 430 300,width=0.50\textwidth}}
  \caption{ Cumulative distribution of image splittings for two
    different quasar surveys. The left panel is for the SDSS
    spectroscopic quasars and the right panel is for SDSS photometric
    quasar sample of \citet{Richards04}. The green (dotted), blue
    (dot-dashed), and red (dashed) curves are the separate
    contributions from the source planes at $z_{\rm s}=1.0$, $2.0$,
    and $4.0$, respectively. The sum of these three curves gives the
    total number of multiply imaged quasars which is the black (solid)
    curve. The thin solid (black) line shows the total number of
    lenses if BCGs are neglected and we ray trace through dark matter
    only.
    \label{fig:N_theta}
  }
\end{figure*}

\begin{figure*}
  \centerline{
    \epsfig{file=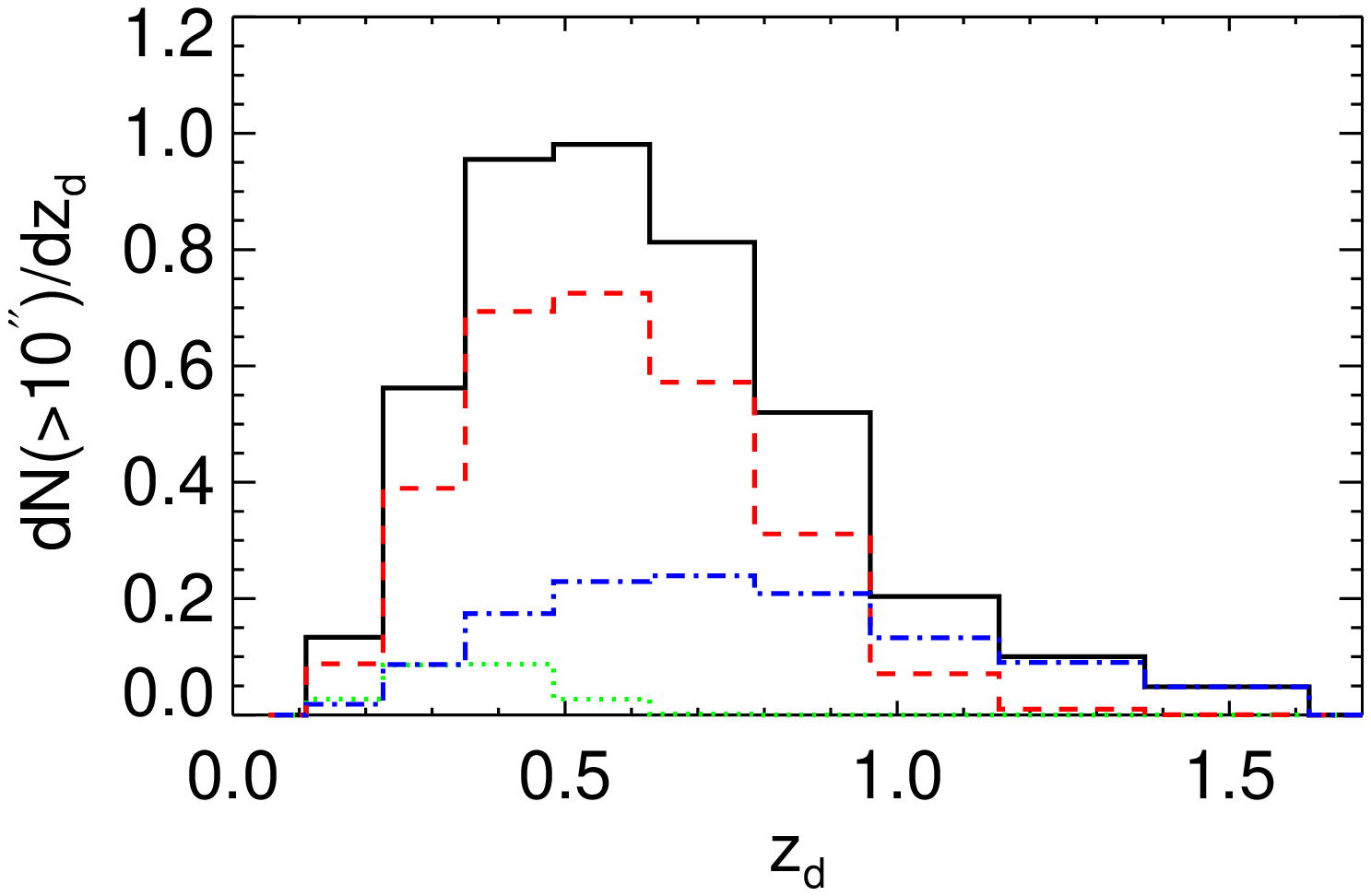,bb=0 10 450 300,width=0.50\textwidth}
    \epsfig{file=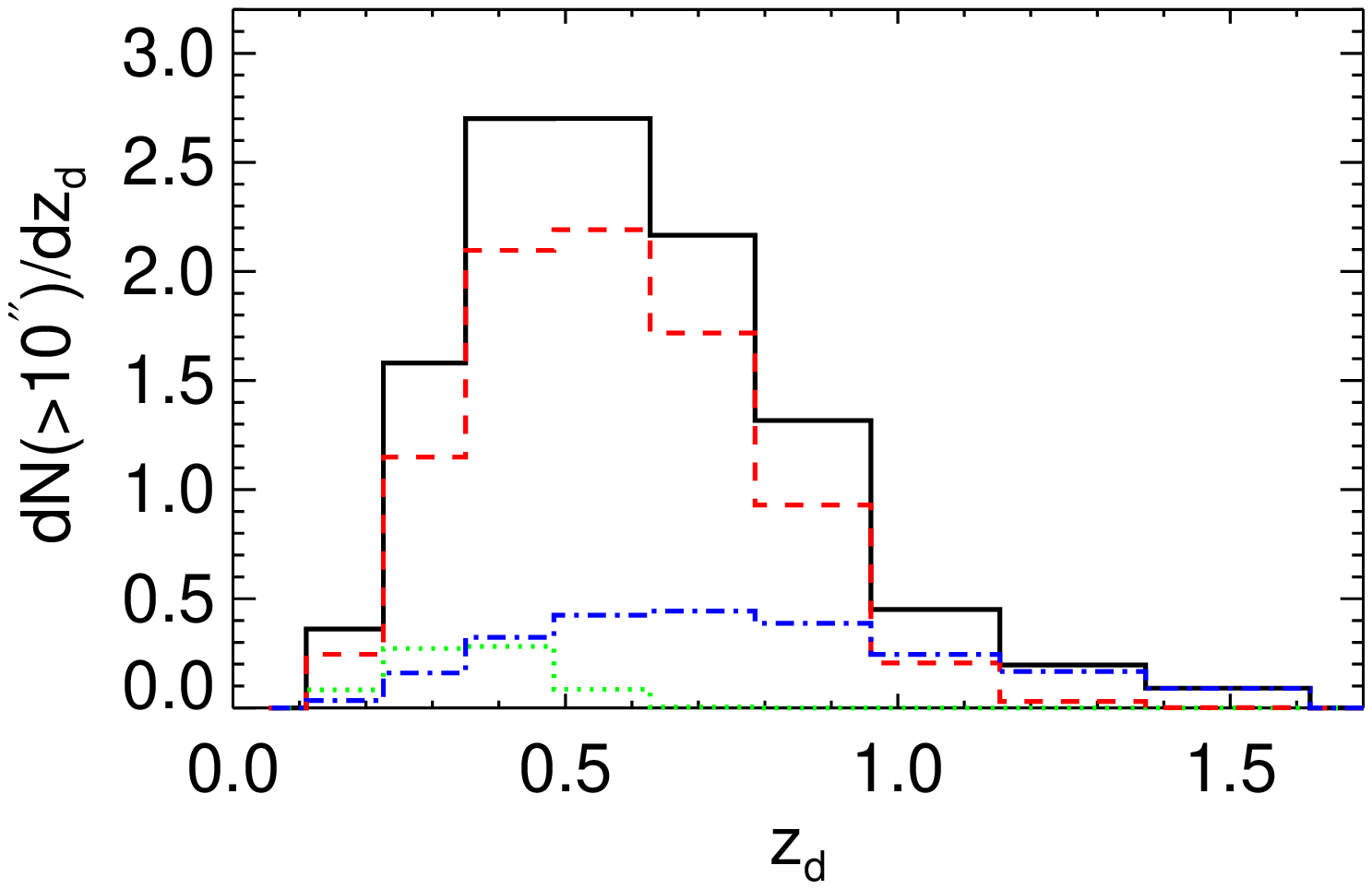,bb=0 10 450 300,width=0.50\textwidth}}
  \caption{ Redshift distribution of cluster lenses with splittings
    $\theta>10\arcsec$ for the spectroscopic quasars (left) and
    photometric quasars (right). The green (dotted), blue
    (dot-dashed), and red (dashed) curves are the separate
    contributions from the source planes at $z_{\rm s}=1.0$, $2.0$, \&
    $4.0$, respectively. The sum of these three curves gives the total
    number of multiply imaged quasars which is the black (solid)
    curve.
    \label{fig:SL_zd}
  }
\end{figure*}

We compute the expected abundance of wide separation multiply imaged quasars 
for two surveys: the SDSS spectroscopic survey \citep{Schneider05} and the
fainter SDSS photometric quasar sample of \citet{Richards04}.

At low redshift $z<2.3$, the quasar luminosity function has been measured by
several groups \citep{Boyle00, Croom04,Richards05}. We use the 
double power law B-band luminosity function \citep{Boyle00}
\be
\Phi(M_{\rm B},z)=\frac{\Phi_{\ast}}{10^{0.4(\beta_{l}+1)[M_{\rm B}-M^{\ast}_{\rm B}(z)]} + 10^{0.4(\beta_{h}+1)[M_{\rm B}-M^{\ast}_{\rm B}(z)]}},
\label{eqn:Boyle}
\ee
where $\beta_l=-1.64$, $\beta_h=-3.43$, and 
$\Phi_{\ast}=360(h\slash 0.50)^3$ Gpc$^{-3}$mag$^{-1}$. 
The evolution of the break  luminosity $M^{\ast}_{\rm B}(z)$ follows
\be
M^\ast_{\rm B}=M^{\ast}_B(0)-2.5(k_1z+k_2z^2),
\label{eqn:mstar}
\ee
with $k_1=1.36$, $k_2=-0.27$,and $M^{\ast}_B(0)=-21.15+5\log h$. 

The quasar luminosity function is poorly constrained between redshifts
$2.3\leq z \leq 3.6$.  \citet{WLlum02} used a simple analytical model 
to fit the double power law luminosity function in eqn.~(\ref{eqn:Boyle}) 
to both the \citet{Fan01} high redshift ($z>3.6$) luminosity function 
and the \citet{Boyle00} low redshift ($z<2.3$) luminosity function, 
using a model for the evolution of the break luminosity proposed by
\citet{MHR99}
\be 
L^\ast_{\rm B}(z)=L^\ast_{\rm B}(0)(1+z)^{(\alpha -1)}\frac{e^{\zeta z}(1+e^{\xi
    z_{\ast}})}{e^{\xi z} + e^{\xi z_{\ast}}}, 
\ee 
where $\alpha=-0.5$ is the slope of a power law spectral energy
distribution assumed for the quasar spectrum $f_{\nu}\sim \nu
^{\alpha}$.  At the bright end they used the \citet{Fan01} slope
$\beta_{\rm h}=2.58$ and assumed a faint end slope $\beta_{\rm
  l}=1.64$. The other parameters of their fit are $\Phi_{\ast}=624$
Gpc$^{-3}$mag$^{-1}$, $M^\ast_{\rm B}(0)=-22.46$, $z_{\ast}=1.60$,
$\xi=2.65$, and $\xi=3.30$. This prescription for the luminosity
function has also been employed in strong lensing studies by
\citet{Quad04} and \citet{OK04}.  For redshifts $z<2.3$ we use the
\citet{Boyle00} expression in eqn.~(\ref{eqn:Boyle}). In the range
$2.3 < z < 3.6$ we linearly interpolate between eqn.~\ref{eqn:Boyle}
and the the \citep{WLlum02} fit, and we use the \citep{WLlum02} fit 
for $z>3.6$.


The luminosity functions we are considering are expressed in terms of
rest frame B-band absolute magnitude, $M_{\rm B}$, whereas the SDSS
spectroscopic and photometric surveys have flux limits in the
$i$-band. We thus require the cross filter K-correction $K_{Bi}(z)$,
between apparent magnitude $i$ and absolute magnitude $B$ \citep[see
  e.g.][]{Hogg02} \be M^i_{\rm B}=i - DM(z) - K_{Bi}(z)
\label{eqn:MB}
\ee 
where $DM(z)$ is the distance modulus. We compute $K_{Bi}(z)$ from
the SDSS composite quasar spectrum of \citet{vanden01} and the
Johnson-B and SDSS $i$ filter curves.

Between redshifts $2.3\leq z \leq 3.6$, quasar surveys which rely on
color selection suffer from incompleteness because quasar colors cross
the stellar locus \citep[see e.g.][]{Richards01,vanden05} in this redshift
range. Because the SDSS spectroscopic and photometric surveys rely on
color selection, we conservatively assume that only 60\% of the
quasars in this interval are recovered, accounting for this
incompleteness.




An area of $\sim 5000$ deg$^2$ is assumed for the SDSS spectroscopic
quasar survey, which has flux limits $i<19.1$, for low redshift
quasars ($z < 2.5$) and $i<20.2$ for high redshift
quasars ($z > 2.5$) \citep{Schneider05}. We assume the
faint photometric quasar survey covers the larger
SDSS imaging area of $\sim 8000$ deg$^2$, with flux limits $i<
21.0$ for $z < 2.5$, and $i<20.5$ for $z > 2.5$ \citep{Richards04}.

Searching for quasar lenses requires requires follow-up observations.
This is the case for the SDSS spectroscopic sample because of
\emph{fiber collisions}\footnote{ The finite size of the optical
  fibers of the SDSS multi object spectrograph imply that only one
  quasar in a close pair with separation $<55\arcsec$ can be will have
  a spectrum in the quasar catalog.}, and also for the faint
photometric sample because spectroscopy is required to confirm that
two quasars in a pair are quasars at the same redshift, rather than a
projected pair of quasars or stellar contaminants. The flux limit of
the follow-up observations is set by the follow-up instrument and is
typically fainter than the flux limit of the parent sample. Thus for
both surveys, we assume a flux limit of $i<21$ for the fainter quasar
in the pair, appropriate for follow up spectroscopy with a 4m class
telescope.

The left panel of Figure~\ref{fig:qso_lum} shows the redshift
distribution of quasars for a flux limit of $i<22.7$, i.e. $2.5$
magnitudes fainter than the SDSS flux limit\footnote{For the sake of
  illustration we have used a single flux limit in the figure although
  both quasar samples have different flux limits for high and low
  redshift.}  of $i<20.2$, corresponding to magnifications $\sim 10$.
The redshift bins used for each source plane are indicated by the
shading.  The right panel of Figure~\ref{fig:qso_lum} shows the
cumulative number magnitude counts of quasars in the redshift bins
about each of the three source planes.

\section{The Number of Quasars Lensed by Clusters}
\label{sec:lens}

The expected cumulative distribution of lens splittings $N(>\theta)$
for the SDSS spectroscopic quasar survey and the SDSS photometric
survey are shown in Figure~\ref{fig:N_theta}. The green (dotted), blue
(dot-dashed), and red (dashed) lines show the individual contributions
from the source planes $z_{\rm s}=1.0$, $z_{\rm s}=2.0$, and $z_{\rm
  s}=4.0$, respectively. The thick solid (black) lines shows the total
number of lenses, which is just the sum from the three individual
source planes. The thin solid (black) line shows the total number of
lenses if the baryons in BCGs are neglected and we ray trace through
dark matter only. 

BCGs increase the number of lenses by $\sim 50\%$, consistent with
previous findings of \citep{Mene03b,DHH04,Shirley04,Hennawi05a}. For
the SDSS spectroscopic quasars, our results are consistent with the
existence of one wide separation lensed quasar in the sample with a
splitting $\theta\sim 14\arcsec$, namely SDSSJ1004+4112. This lens was
discovered in an effective area of $\sim 2500$ deg$^2$
\citep{Quad03,Quad04}, which is roughly half of that used in the left
panel of Figure~\ref{fig:N_theta}, so we would predict $1-2$ such
lenses in a quasar sample of this size.  Our prediction of $\sim 4$
quasar lenses with splittings $\theta>10\arcsec$, suggests that
several wide separation quasar lenses may still be lurking in the SDSS
spectroscopic survey.  For the faint photometric quasars, our
calculation predicts $\sim 12$ multiply imaged quasars with splittings
$\theta>10\arcsec$. It is expected that $\sim 2$ lenses will have a
separation $\theta>30\arcsec$, and $\sim 2$ wide separation lenses
will be at $z_{\rm s}\sim 4$.

At first glance, it seems unexpected that the number of quasar lenses
in the photometric sample is only a factor of $\sim 3$ higher than the
spectroscopic sample, considering that the surface density of quasars
increases by a factor of $\sim 6$ and the area is increased by a
factor of $\sim 2$ \citep{Richards04}.  However, recall that the
flux limit of the faintest image in the lens, which is identified from
follow-up observations, is the same for both samples, namely $i<21$.
Most of the lensing probability is in the source bins $z_{\rm s}=2.0$
and $z_{\rm s}=4.0$, and the SDSS spectroscopic survey flux limit for
these bins is already $i < 20.2$, compared to $i<21.0$ for $z_{\rm
  s}=2.0$ and $i<20.5$ for $z_{\rm s}=4.0$ in the photometric
sample. Thus the fainter flux limits of the photometric sample do not
increase the number of lenses as much as naively expected.

In Figure~\ref{fig:SL_zd} we show the redshift distribution of the
cluster lenses with $\theta>10\arcsec$ for both surveys. The lens
redshift distribution peaks around $z_{\rm d}=0.5$. Although the
efficiency for strong lensing peaks at half distance to the source
plane, which is at $z_{\rm d}=0.75$ for the dominant source plane
$z_{\rm s}=2.0$, the distribution is skewed toward the
observer by the cluster mass function. Our results for the redshift
distribution of the SDSS cluster lenses is consistent with the redshift, 
$z=0.68$, of the cluster multiply imaging SDSSJ1004+4112. 

\ \\

\section{Summary and Conclusions}
\label{sec:conc}
We consider the following to be the primary conclusions of this study:

\begin{enumerate}

\item
Our ray tracing calculation through clusters from an N-body simulation
of the $\Lambda$CDM cosmology predicts $\sim 4$ quasar lenses with
splittings $\theta>10\arcsec$ in the SDSS spectroscopic quasar sample,
consistent with the recent discovery of the wide separation lens
SDSSJ1004+4112 which has $\theta=14\arcsec.6$.

\item

Our study predicts $\sim 12$ multiply imaged quasars with splittings
$\theta>10\arcsec$ should be found in the faint photometric quasar
sample of \citet{Richards04}. Of these, $\sim 2$ should be lenses
with $\theta>30\arcsec$ and $\sim 2$ will be at high redshift
$z_{\rm s}\sim 4$.

\item

BCGs increase the number of multiply imaged quasar lenses by $\sim
50\%$, consistent with previous findings in the context of giant arcs 
\citep{Mene03b,DHH04,Shirley04,Hennawi05a}.

\end{enumerate}

The search for wide separation multiply imaged quasar lenses is
currently underway in both the SDSS spectroscopic
\citep{Quad03,Quad04,Oguri05,Hennawi05b} and photometric samples
\citep{Richards04,Hennawi05b}. Looking forward, wide field synoptic
surveys like Pan-STARRS and the LSST will be able to combine
variability and color selection techniques to detect as many as $10^8$
quasar down to $i\lesssim 27$ (Richard Green, private communication
2004). Crudely extrapolating from the lensing rate computed here,
these future surveys are likely to contain $\gtrsim 300$ quasars
multiply imaged by clusters. A comparison of the abundance of these
new lenses to theoretical predictions as well as detailed modeling
of each system, will provide a powerful test of the $\Lambda$CDM 
cosmological model.

\acknowledgments

We are grateful to Jerry Ostriker for reading an early version of this
manuscript and providing helpful comments. We also wish to thank Jim
Gunn, Richard Green, Donald Schneider, and Gordon Richards for helpful
discussions. JFH would like to thank his thesis advisors David Spergel
and Michael Strauss for advice and guidance during his time in
Princeton, where this work was started.  For part of this study JFH was
supported by Proctor Graduate fellowship at Princeton University and
by a generous gift from the Paul \& Daisy Soros Fellowship for New
Americans. The program is not responsible for the views expressed.
JFH and ND are supported by NASA through Hubble Fellowship grants \#
01172.01-A and 01148.01-A respectively, awarded by the Space Telescope
Science Institute, which is operated by the Association of
Universities for Research in Astronomy, Inc., for NASA, under contract
NAS 5-26555.  Computer time was provided by the National Computational
Science Alliance under program \#MCA04N002P, and some computations were
performed on the NSF Terascale Computing System at the Pittsburgh
Supercomputing Center. The ray tracing simulations used computational
facilities at Princeton supported by NSF grant AST-0216105.


\begin{thebibliography}
\frenchspacing


\bibitem[Bartelmann, Steinmetz, \& Weiss(1995)]{BSW95}
Bartelmann, M., Steinmetz, M., \& Weiss, A.\ 1995, \aap, 297, 1 

\bibitem[Bartelmann et al.(1998)]{Bart98} Bartelmann, M., 
Huss, A., Colberg, J.~M., Jenkins, A., \& Pearce, F.~R.\ 1998, \aap, 330, 1 

\bibitem[Bartelmann et al.(2003)]{Bart03} Bartelmann, M., 
Meneghetti, M., Perrotta, F., Baccigalupi, C., \& Moscardini, L.\ 2003, 
\aap, 409, 449 

\bibitem[Bode \& Ostriker(2003)]{BO03} Bode, P.~\& Ostriker, 
J.~P.\ 2003, \apjs, 145, 1

\bibitem[Boyle et al.(2000)]{Boyle00} Boyle, B.~J., Shanks, T., 
Croom, S.~M., Smith, R.~J., Miller, L., Loaring, N., \& Heymans, C.\ 2000, 
\mnras, 317, 1014

\bibitem[Chen(2004)]{Chen04} Chen, D.-M.\ 2004, \aap, 418, 387 

\bibitem[Croom et al.(2004)]{Croom04} Croom, S.~M., Smith, 
R.~J., Boyle, B.~J., Shanks, T., Miller, L., Outram, P.~J., \& Loaring, 
N.~S.\ 2004, \mnras, 349, 1397 

\bibitem[Dalal, Holder, \& Hennawi(2004)]{DHH04} Dalal, N., 
Holder, G., \& Hennawi, J.~F.\ 2004, \apj, 609, 50

\bibitem[Dalal et al.(2005)]{DHB05} Dalal, N., Hennawi, 
J.~F., \& Bode, P.\ 2005, \apj, 622, 99

\bibitem[Davis \etal(1985)]{Davis85} 
Davis, M., Efstathiou, G., Frenk, C.~S., \& White, S.~D.~M.\ 1985, \apj, 
292, 371

\bibitem[Edge \& Stewart(1991)]{ES91} Edge, A.~C.~\& 
Stewart, G.~C.\ 1991, \mnras, 252, 428

\bibitem[Fan et al.(2001)]{Fan01} Fan, X., et al.\ 2001, \aj, 
121, 54

\bibitem[Hennawi et al.(2005a)]{Hennawi05a} Hennawi. J.~F., Dalal, N.,
  Bode, P., Ostriker, J.~P.  \etal 2005, ArXiv Astrophysics e-prints, 
  arXiv:astro-ph/0506171 

\bibitem[Hennawi et al.(2005b)]{Hennawi05b} Hennawi, J.~F., et al.\ 
2005, ArXiv Astrophysics e-prints, arXiv:astro-ph/0504535

\bibitem[Hogg \etal(2002)]{Hogg02} Hogg,
  D.~W., Baldry, I.~K., Blanton, M.~R., Eisenstein, D.~J. \ 2002,
  ArXiv Astrophysics e-prints, astro-ph/0210394

\bibitem[Ho \& White(2004)]{Shirley04} Ho, S., \& White, M.\ 
  2004, ArXiv Astrophysics e-prints, astro-ph/0408245

\bibitem[Huterer \& Ma(2004)]{HM04} Huterer, D.~\& Ma, C.\ 
2004, \apjl, 600, L7

\bibitem[Inada et al.(2003)]{Quad03} Inada, N., et al.\ 2003, 
\nat, 426, 810 

\bibitem[Inada et al.(2005)]{Inada05} Inada, N., et al.\ 2005, 
ArXiv Astrophysics e-prints, arXiv:astro-ph/0503310

\bibitem[Keeton \& Madau(2001)]{KM01} Keeton, C.~R.~\& 
Madau, P.\ 2001, \apjl, 549, L25 


\bibitem[Kochanek(1996)]{Koch96} Kochanek, C.~S.\ 1996, \apj, 
466, 638 
\bibitem[Kochanek(2003)]{Koch03} Kochanek, C.~S.\ 2003, \apj, 
583, 49 

\bibitem[Kochanek(2004)]{Castles} Kochanek, C. S. \etal CASTLES
  Survey. at $<$http://cfa-www.harvard.edu/castles/$>$ (2003).

\bibitem[Kochanek \& Schechter(2004)]{KS04} Kochanek, 
C.~S.~\& Schechter, P.~L.\ 2004, Measuring and Modeling the Universe, 117

\bibitem[Kuhlen, Keeton, \& Madau(2004)]{KKM04} Kuhlen, M., 
Keeton, C.~R., \& Madau, P.\ 2004, \apj, 601, 104

\bibitem[Li \& Ostriker(2002)]{LO02} Li, L.~\& Ostriker, 
J.~P.\ 2002, \apj, 566, 652 

\bibitem[Li \& Ostriker(2003)]{LO03} Li, L.~\& Ostriker, 
J.~P.\ 2003, \apj, 595, 603 

\bibitem[Lopes \& Miller(2004)]{LM04} Lopes, A.~M.~\& 
Miller, L.\ 2004, \mnras, 348, 519

\bibitem[Madau, Haardt, \& Rees(1999)]{MHR99} Madau, P., 
Haardt, F., \& Rees, M.~J.\ 1999, \apj, 514, 648 

\bibitem[Maoz \etal(1997)]{Maoz97} Maoz, D., 
Rix, H., Gal-Yam, A., \& Gould, A.\ 1997, \apj, 486, 75

\bibitem[Marble et al.(2004)]{Marble04} Marble, A.~R., Impey, 
C.~D., Miller, L., Clewley, L., Edmondson, E., \& Lopes, A.~M.\ 2004, 
American Astronomical Society Meeting Abstracts, 205,  

\bibitem[Meneghetti, Bartelmann, \& Moscardini(2003a)]{Mene03a} 
Meneghetti, M., Bartelmann, M., \& Moscardini, L.\ 2003, \mnras, 340, 105 

\bibitem[Meneghetti, Bartelmann, \& Moscardini(2003b)]{Mene03b}
Meneghetti, M., Bartelmann, M., \& Moscardini, L.\ 2003, \mnras, 346, 67 


\bibitem[Meneghetti et al.(2004)]{Mene04} Meneghetti, M., 
  Bartelmann, M., Dolag, K., Moscardini, L., Perrotta, F., Baccigalupi, C., 
  \& Tormen, G.\ 2004, ArXiv Astrophysics e-prints, astro-ph/0405070 

\bibitem[Miller et al.(2004)]{Miller04} Miller, L., Lopes, 
A.~M., Smith, R.~J., Croom, S.~M., Boyle, B.~J., Shanks, T., \& Outram, P.\ 
2004, \mnras, 348, 395 

\bibitem[Oegerle \& Hoessel(1991)]{OH91} Oegerle, W.~R.~\& 
Hoessel, J.~G.\ 1991, \apj, 375, 15

\bibitem[Ofek et al.(2001)]{Ofek01} Ofek, E.~O., Maoz, D., 
Prada, F., Kolatt, T., \& Rix, H.\ 2001, \mnras, 324, 463

\bibitem[Oguri(2003)]{Oguri03} Oguri, M.\ 2003, \mnras, 339, 
L23 

\bibitem[Oguri et al.(2004)]{Quad04} Oguri, M., et al.\ 2004, 
\apj, 605, 78 

\bibitem[Oguri \& Keeton(2004)]{OK04} Oguri, M.~\& Keeton, 
C.~R.\ 2004, ArXiv Astrophysics e-prints, astro-ph/0403633

\bibitem[Oguri et al.(2005)]{Oguri05} Oguri, M., et al.\ 2005, 
\apj, 622, 106

\bibitem[Phillips et al.(2001)]{Phillips01} Phillips, P.~M., et 
al.\ 2001, \mnras, 328, 1001 

\bibitem[Richards et al.(2001)]{Richards01} Richards, G.~T., et 
al.\ 2001, \aj, 121, 2308 

\bibitem[Richards et al.(2004)]{Richards04} Richards, G.~T., et 
al.\ 2004, \apjs, 155, 257

\bibitem[Richards et al.(2005)]{Richards05} Richards, G.~T., et 
al.\ 2005, ArXiv Astrophysics e-prints, arXiv:astro-ph/0504300 

\bibitem[Rusin(2002)]{Rusin02} Rusin, D.\ 2002, \apj, 572, 705

\bibitem[Sarbu, Rusin, \& Ma(2001)]{SRM01} Sarbu, N., Rusin, 
D., \& Ma, C.\ 2001, \apjl, 561, L147 

\bibitem[Schneider et al.(2005)]{Schneider05} Schneider, D.~P., et 
al.\ 2005, ArXiv Astrophysics e-prints, arXiv:astro-ph/0503679

\bibitem[Spergel et al.(2003)]{Spergel03} Spergel, D.~N., et al.\ 
2003, \apjs, 148, 175 


\bibitem[Vanden Berk et al.(2001)]{vanden01} Vanden Berk, D.~E., 
et al.\ 2001, \aj, 122, 549

\bibitem[Vanden Berk et al.(2005)]{vanden05} Vanden Berk, D.~E., 
et al.\ 2005, \aj, 129, 2047

\bibitem[Wambsganss et al.(2004a)]{Wamb04a} Wambsganss, J., 
Bode, P., \& Ostriker, J.~P.\ 2004, \apjl, 606, L93 

\bibitem[Wambsganss et al.(2004b)]{Wamb04b} Wambsganss, J., 
Bode, P., \& Ostriker, J.~P.\ 2004, ArXiv Astrophysics e-prints, 
astro-ph/0405147 

\bibitem[Wyithe, Turner, \& Spergel(2001)]{WTS01} Wyithe, 
J.~S.~B., Turner, E.~L., \& Spergel, D.~N.\ 2001, \apj, 555, 504 

\bibitem[Wyithe \& Loeb(2002)]{WLlum02} Wyithe, J.~S.~B.~\& 
Loeb, A.\ 2002, \apj, 577, 57 

\bibitem[York et al.(2000)]{York00} York, D.~G., et al.\ 2000, 
\aj, 120, 1579 

\bibitem[Zhdanov \& Surdej(2001)]{ZS01} Zhdanov, V.~I.~\& 
Surdej, J.\ 2001, \aap, 372, 1

\end{thebibliography}
\end{document}